\documentclass{aa}  

\usepackage{graphicx,txfonts,lscape,natbib}
\bibpunct{(}{)}{;}{a}{}{,} % to follow the A&A style

\begin{document}

   \title{Lithium in the Hyades L5 brown dwarf 2MASS\,J04183483$+$2131275\thanks{Based on observations made with the Gran Telescopio de Canarias (GTC) installed in the Spanish Observatorio del Roque de los Muchachos of the Instituto de Astrof\'isica de Canarias, in the island of La Palma (programme GTC77-16B)}}

%   \subtitle{}

   \author{N.\ Lodieu \inst{1,2}
        \and
       R.\ Rebolo \inst{1,2,3}
       \and
       A.\ P\'erez-Garrido \inst{4}
        }

   \institute{Instituto de Astrof\'isica de Canarias (IAC), Calle V\'ia L\'actea s/n, E-38200 La Laguna, Tenerife, Spain \\
       \email{nlodieu,rrl@iac.es}
       \and
       Departamento de Astrof\'isica, Universidad de La Laguna (ULL), E-38206 La Laguna, Tenerife, Spain
       \and
       Consejo Superior de Investigaciones Cient\'ificas, CSIC, Spain
       \and
       Dpto.\ F\'\i sica Aplicada, Universidad Polit\'ecnica de Cartagena, E-30202  Cartagena, Murcia, Spain
       }

   \date{Received \today{}; accepted (date)}

% \abstract{}{}{}{}{} 
% 5 {} token are mandatory
 
  \abstract
   {}
   {From the luminosity, effective temperature, and age of the Hyades brown dwarf 2MASS\,J04183483$+$2131275
    (2M0418), sub-stellar evolutionary models predict a mass in the range 
    39$-$55 Jupiter masses ($M_{\rm Jup}$)
    which is insufficient to produce any substantial lithium burning except for the very upper range 
    $>$53 $M_{\rm Jup}$. Our goal is to measure the abundance of lithium in this object, test the consistency 
    between models and observations and refine constraints on the mass and age of the object.
    }
   {We used the 10.4-m Gran Telescopio Canarias (GTC) with its low-dispersion optical spectrograph to obtain 
    ten spectra of 2277s each covering the range 6300--10300\,\AA{} with a resolving power of R\,$\sim$\,500\@.
   }
   {In the individual spectra, which span several months, we detect persistent unresolved H$\alpha$ in emission 
    with pseudo equivalent widths (pEW) in the range 45--150\,\AA{} and absorption lines of various alkalis 
    with the typical strengths found in objects of L5 spectral type. The lithium resonance line at 6707.8\,\AA{}
    is detected with pEW of 18$\pm$4\,\AA{} in 2M0418 (L5).
    }
   {We determine a lithium abundance of $\log$\,N(Li)\,=\,3.0$\pm$0.4 dex consistent with a minimum preservation
    of 90\% of this element which confirms 2M0418 as a brown dwarf with a maximum mass of 
    52 $M_{\rm Jup}$. We infer a maximum age for the Hyades of 775 Myr from a comparison with 
    the BHAC15 models. 
    Combining recent results from the literature with our study, we constrain the mass of 2M0418
    to 45--52 $M_{\rm Jup}$ and the age of the cluster to 580--775 Myr (1$\sigma$) based on the
    lithium depletion boundary method.
   }  


  \authorrunning{Lodieu, Rebolo \& P\'erez-Garrido.}
  \titlerunning{Lithium in a L5 member of the Hyades}

   \maketitle
%
%________________________________________________________________

%
%%%%%%%%%%%%%%%%%%%%%%%%%%%%%%
%%%%%  Introduction  %%%%%
%%%%%%%%%%%%%%%%%%%%%%%%%%%%%%
%
\section{Introduction}
\label{HyadesL5:intro}

Among the light elements, $^7$Li plays a special role as a discriminant of sub-stellar nature. In stellar interiors 
the $^7$Li nuclei are burnt via collisions with protons at temperatures above 2.5 x 10$^6$ K, but solar metallicity 
brown dwarfs (BDs) with masses below 50 $M_{\rm Jup}$ cannot reach this temperature 
\citep[e.g.][]{baraffe03,magazzu93}.
Lithium is preserved from destruction in these BDs and should be present in their atmospheres with an 
abundance close to the cosmic value \citep{rebolo92}. Strong lithium features have been seen in a large variety 
of BDs \citep{rebolo96,rebolo98,kirkpatrick08,faherty14a,lodieu15b}.

Objects more massive than 50 $M_{\rm Jup}$ progressively burn lithium as they age. In such objects 
the lithium abundance is a very sensitive function of mass and age, parameters rather difficult to determine for 
most isolated BDs. While the detection and abundance determination of lithium provides an important 
indication of sub-stellarity, by itself it has a limited capacity to restrict these two parameters in such 
objects \citep{baraffe98,siess00,baraffe15,feiden15a}. 

However, the coevality of the members of a stellar cluster offer a remarkable opportunity to test the predictions 
of lithium destruction as a function of mass for BDs. The Li-mass (Li-effective temperature or 
Li-luminosity) depletion curves are very sensitive to the age of a cluster and systematic observations of lithium 
in the sub-stellar domain of a cluster can constrain its age 
\citep{stauffer98,jeffries03,oliveira03,barrado04b,manzi08,dobbie10}. 
In addition, an empirical determination of the Li-mass relation provides physical insight on the parameters 
which govern the maximum temperature reached in the interior of BDs, and in particular on the equation 
of state.

%
%%%%%%%%%%%%%%%%%%%%%%%%%%%%%%%%%%%%%%%%%
%%%%% Figure: Final Spectrum  %%%%%
%%%%%%%%%%%%%%%%%%%%%%%%%%%%%%%%%%%%%%%%%
%
\begin{figure*}
 \centering
  \includegraphics[width=0.63\linewidth, angle=0]{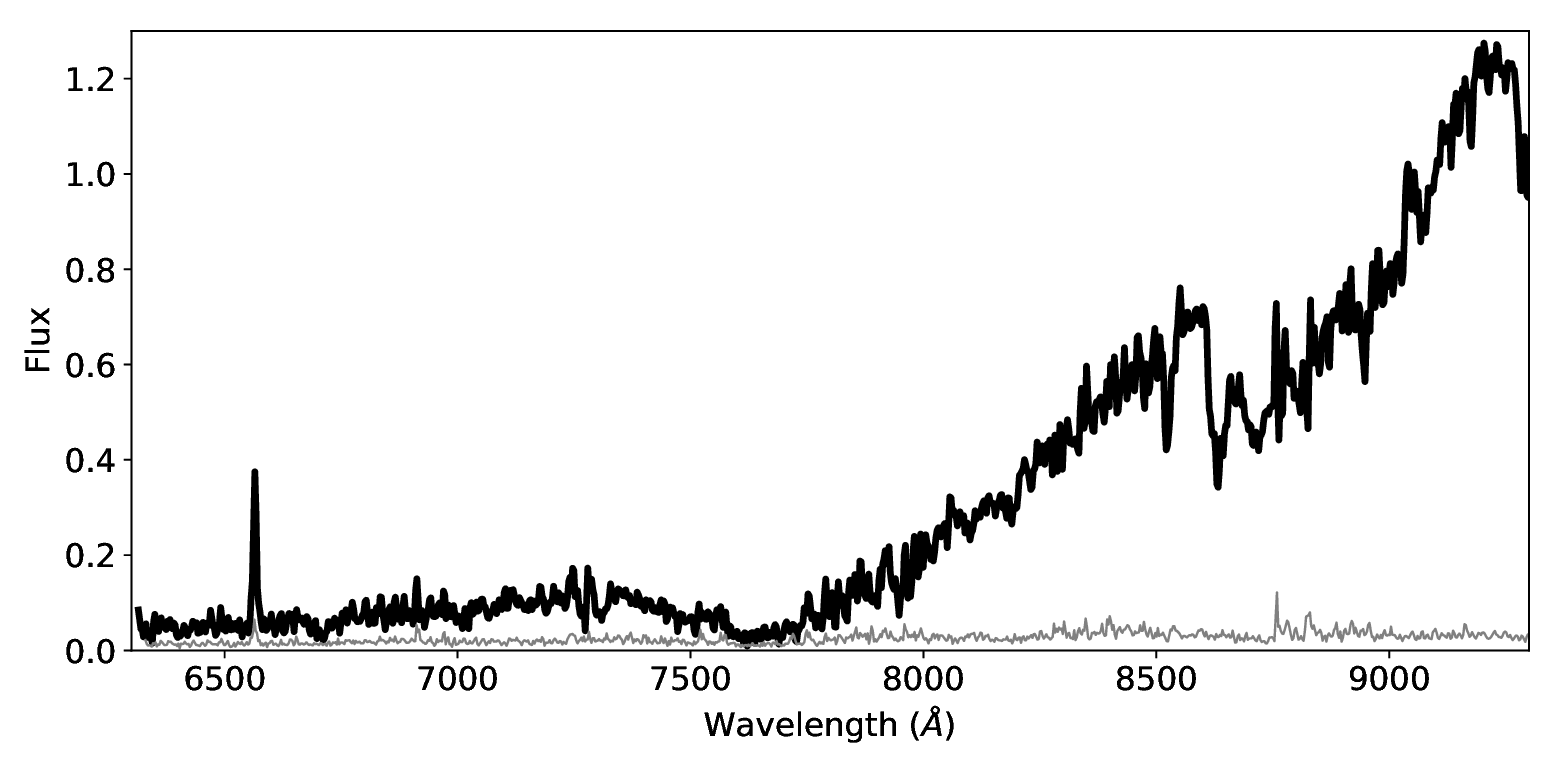}
  \includegraphics[width=0.31\linewidth, angle=0]{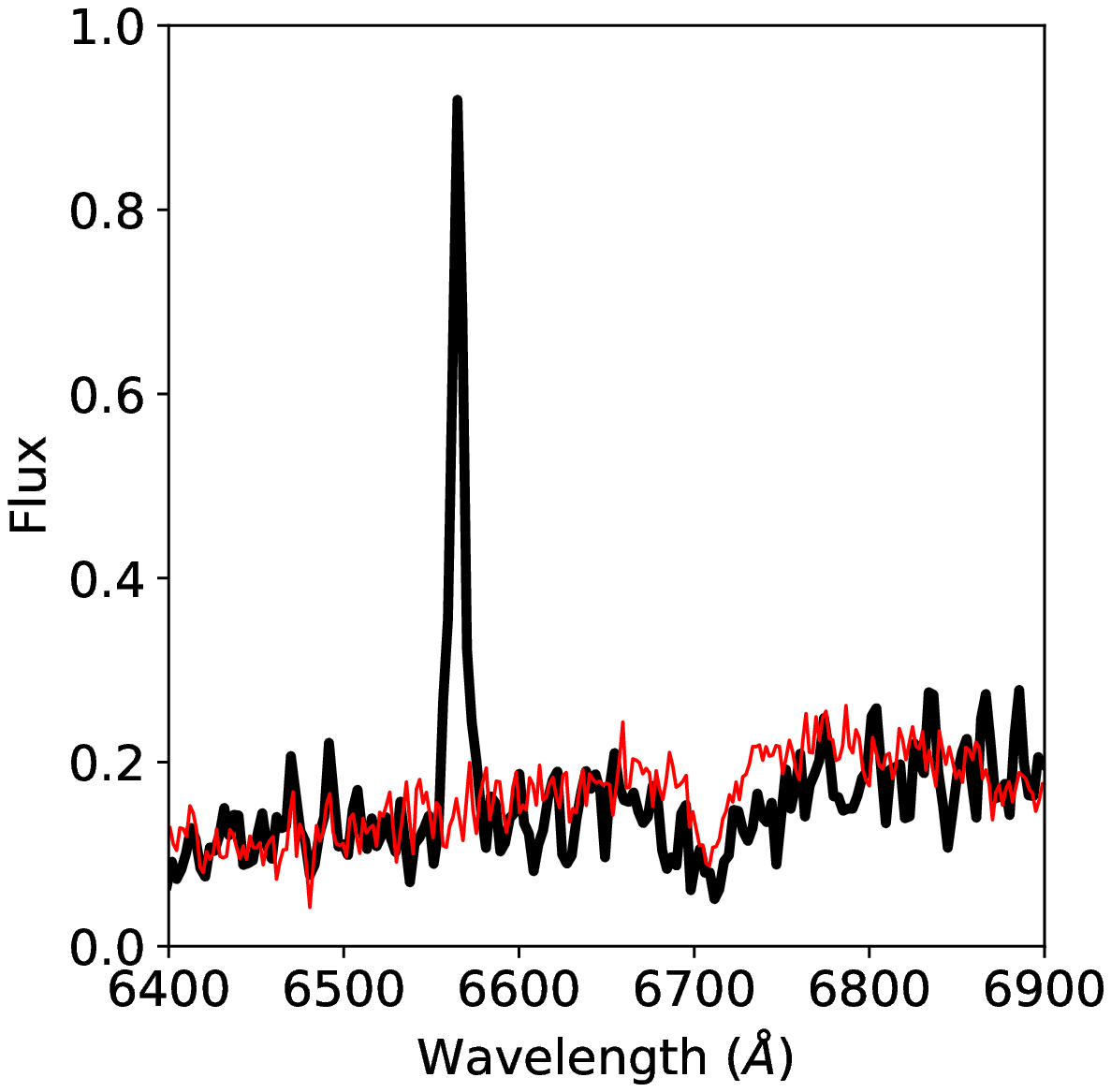}
  \caption{Full GTC/OSIRIS spectrum of the Hyades L5 dwarf 2M0418 (black) calibrated
in flux with the standard deviation from the ten individual optical spectra (grey). 
The right-hand plot shows a zoom on H$\alpha$ in emission and lithium in absorption.
Overplotted in red is the optical spectrum of the field L5V dwarf 2MASS\,J12392727$+$5515371 
which exhibit lithium in absorption \citep{kirkpatrick99}.
}
  \label{fig_HyadesL5:Spectrum}
\end{figure*}

The luminosity and effective temperature (T$_{\rm eff}$) of the L5 member of the Hyades 
2MASS\,J04183483$+$2131275 (hereafter 2M0418) \citep{perez-garrido17}, indicates a mass range
of 39--55 $M_{\rm Jup}$ comparing its $J-K$, $J-W2$, and $W1-W2$ colours with the AMES-Dusty 
evolutionary models \citep{allard01,chabrier00c}, hence a significant preservation of the original 
lithium content is expected.
However, the conversion from BD parameters (luminosity and T$_{\rm eff}$) to masses are subject 
to various uncertainties: if the mass were only 10\% higher -\ 60 $M_{\rm Jup}$ - a significant 
lithium  destruction (90\%) could have taken place by the age of the Hyades. This object offers an opportunity 
to observe a relatively old BD ($>$\,500\,Myr) with fully 
preserved lithium, or alternatively, if depletion is already undergoing, to impose very narrow constraints on 
its mass from theoretical lithium depletion evolutionary curves. In the case of full preservation the lithium 
resonance line is expected to be rather strong in L dwarfs \citep{pavlenko00a,kirkpatrick08}
and detectable in such faint object (R\,$\sim$\,23 mag). Lithium has been detected even in 
cooler objects, such as the very nearby binary BD Luhman\,16AB \citep{faherty14a,lodieu15b} showing that 
the formation of Li-based molecules \citep[e.g.\ LiH, LiCl;][]{lodders99} is not a problem for detecting 
the Li resonance doublet at these temperatures. 

In this manuscript, we present spectroscopic observations of 2M0418 (Sect.\ \ref{HyadesL5:spec_obs})
and derive its physical parameters (Sect.\ \ref{HyadesL5:physical_param}). We discuss the strong lithium
feature at 6707.8\,\AA{}, which according to spectral synthesis models is consistent with a cosmic
abundance of lithium and therefore, with full (or close to full) preservation in this BD\@.
We also present the detection of H$\alpha$ in emission Sect.\ \ref{HyadesL5:Halpha}.
Finally, we derive the age of the Hyades combining our detection of lithium and the presence
of lithium in higher mass BDs (Sect.\ \ref{HyadesL5:age_Hyades}).
For older clusters, the lithium depletion boundary moves to later-type and fainter members,
which explains the age of the cluster has only been recently addressed by \citet{martin18a}
and this work.

%
%%%%%%%%%%%%%%%%%%%%%%%%%%%%%%%%%%%%%%%%%%
%%%%% Spectroscopic Observations %%%%%
%%%%%%%%%%%%%%%%%%%%%%%%%%%%%%%%%%%%%%%%%%
%
\section{Spectroscopic observations}
\label{HyadesL5:spec_obs}

We collected ten optical spectra with OSIRIS \citep[Optical System for Imaging and low-intermediate Resolution 
Integrated Spectroscopy;][]{cepa00} on the 10.4-m GTC at Observatorio del Roque de los 
Muchachos (La Palma, Spain) over several nights between 23 January and 28 February 2017 as part of programme 
GTC77-16B (PI P\'erez-Garrido). 

OSIRIS is equipped with two 2048$\times$4096 Marconi CCD42-82 detectors offering a field of view of 
approximately 7$\times$7 arcmin$^2$ with a binned pixel scale of 0.25 arcsec.
We employed the R1000R grating with a slit of 1.23 arcsec, yielding a spectral resolution of approximately
500 over the 6300--10300\,\AA{} range.

We obtained five separate observing blocks on five distinct nights: 23 January, 2, 16, 26, and 28 February 2017\@.
Each block was made of two individual spectra of 2277s nodded along the slit to improve the sky subtraction,
yielding a total exposure of 6.325 hours.
Observations were obtained under seeing in the 0.7--1.2 arcsec range, clear conditions and grey or dark time.

We reduced the optical spectra under the IRAF environment \citep{tody86,tody93} in a standard
manner. First, we median-combined the bias and flat-fields taken during the afternoon. We
subtracted the mean bias from the raw spectrum of the target and divided by the normalised flat
field. We subtracted the first position from the second one of each observing block and extracted
optimally the individual spectra by choosing manually aperture and sky regions\footnote{We used the
routine {\tt{apsum}} in IRAF \citep{tody86,tody93}.}
We calibrated each spectrum in wavelength with the arc Xe$+$Ne$+$HgAr lamps observed during the preceding
afternoon. We combined all ten spectra with a median filter and estimate the error at each wavelength
from the standard deviation of the individual spectra (Fig.\ \ref{fig_HyadesL5:Spectrum}).

We computed various spectral indices defined by \citet{kirkpatrick99} and \citet{cruz09} for this
intermediate-age mid-L dwarf. We report the values in Table \ref{tab_HyadesL5:spec_indices}.
{We observe that these values are in agreement with intermediate-age L dwarfs comparing with 
the ranges reported by \citet{cruz09}, consistent with the age of the Hyades cluster. The cesium line at 
8521\AA{} appears also weaker than in field L5 dwarfs, supporting the intermediate-age of the cluster 
because this line is gravity-sensitive. The other spectral indices available in the literature are affected 
by low signal-to-noise regions of our combined spectrum, so we do not report them in this work.

%
%%%%%%%%%%%%%%%%%%%%%%%%%%%%%%%%%%%%%%%%%%%%%%%%%%
%%%%% Table: Spectral indices %%%%%
%%%%%%%%%%%%%%%%%%%%%%%%%%%%%%%%%%%%%%%%%%%%%%%%%%
%
\begin{table}
%\scriptsize
\centering
\caption{Spectral indices for the intermediate-age mid-L dwarf 2M0418.}
\begin{tabular}{@{\hspace{0mm}}c @{\hspace{1mm}}c @{\hspace{1mm}}c @{\hspace{1mm}}c @{\hspace{1mm}}c @{\hspace{1mm}}c @{\hspace{1mm}}c @{\hspace{1mm}}c @{\hspace{1mm}}c @{\hspace{1mm}}c @{\hspace{1mm}}c @{\hspace{1mm}}c @{\hspace{1mm}}c@{\hspace{0mm}}}
\hline
\hline
SpT &  TiO-a & TiO5  & CrH-a  &  Rb-a  &   Na-a  &  Na-b  & FeH-a \cr
L5  &  1.116 & 1.210 & 1.726  &  1.489 &   1.009 &  1.024 & 1.214 \cr
\hline
\label{tab_HyadesL5:spec_indices}
\end{tabular}
\end{table}
%

%
%%%%%%%%%%%%%%%%%%%%%%%%%%%%%%%%%%%
%%%%% Physical parameters  %%%%%
%%%%%%%%%%%%%%%%%%%%%%%%%%%%%%%%%%%
%
\section{Effective temperature and luminosity}
\label{HyadesL5:physical_param}

We computed the distance of 2M0418 using the latest version of the BANYAN $\Sigma$ algorithm 
\citep{gagne18a} 
which provides a kinematic distance estimated based on the XYZ and UVW distribution of Hyades members,
yielding a mean value of 
40.6$\pm$2.7 pc\footnote{http://www.exoplanetes.umontreal.ca/banyan/banyansigma.php?radeg=64.645125\&pmra=124\&pmdec=-53\&hrv=38\&!plx=\&submit=Submit\&dec=21.524306\&epmra=7\&epmdec=6\&ehrv=2.9\&eplx=\&targetname=0418}.

We estimated the T$_{\rm eff}$ of 2M0418 using the most recent polynomials based on the spectral 
types and absolute $H$-band magnitudes of field M6--T9 \citep{filippazzo15}.
We derived consistent mean T$_{\rm eff}$ of 1581$\pm$113 and 1516$\pm$29\,K based on the L5 spectral type
and the UKIDSS $H$-band magnitude \citep{lawrence07} corrected for the distance (40.6$\pm$2.7 pc).
The uncertainties on the T$_{\rm eff}$ does include the dispersion from the polynomials of
\citet{filippazzo15} but not the uncertainty on the spectral type determination.
To summarise, we assign a mean T$_{\rm eff}$ of 1581$\pm$113K to 2M0418\@.
We note that we might be over-estimating the T$_{\rm eff}$ of this L5 member of the 
Hyades by up to about 50\,K because it has an age lower than the field dwarfs in the 
sample of \citep{filippazzo15}. 
The young sample in that study indicates a temperature 100\,K lower for L dwarfs with ages between 10 and 120 Myr.
We plot this measurement in Fig.\ \ref{fig_HyadesL5:Li_vs_Teff} where we display the evolution of
the Li pEW in Hyades solar-type stars \citep{thorburn93,cummings17a}, the upper limits on the
lithium features in the optical spectra of K and M low-mass stars \citep{stauffer97b} with 
T$_{\rm eff}$ taken from \citet{rajpurohit13}, and the mean value for Hyades L3.5 dwarfs \citep{martin18a}.

In Table 2 of \citet{perez-garrido17}, we derived a bolometric luminosity of 2.80$\times$10$^{29}$ erg/s,
translating in $\log$(L$_{\rm bol}$/L$_{\odot}$) of $-$4.14 dex.
We re-calculated the value of $\log$(L$_{\rm bol}$/L$_{\odot}$) using the UKIDSS $K$-band magnitude
(15.230$\pm$0.022), the distance from BANYAN, and the
$K$-band bolometric correction for a L5 dwarf (3.30$\pm$0.05 mag) from \citet{filippazzo15}, yielding 
a slightly lower value of $-$4.30$\pm$0.07 dex (Table \ref{tab_HyadesL5:param}).

%
%%%%%%%%%%%%%%%%%%%%%%%%%%%%%%%%%%%%%%%%%%%%%%%%%%
%%%%% Table: physical parameters %%%%%
%%%%%%%%%%%%%%%%%%%%%%%%%%%%%%%%%%%%%%%%%%%%%%%%%%
%
\begin{table}
\centering
\caption{Physical parameters of 2M0418 derived in this study.
}
\begin{tabular}{c c}
\hline
\hline
SpT & L5.0$\pm$0.5 \\
$\log$(L$_{\rm bol}$/L$_{\odot}$) & $-$4.30$\pm$0.07 dex \\
T$_{\rm eff}$ &  1581$\pm$113K  \\
pEW (Li) &  18$\pm$4\,\AA{} \\
pEW (H$\alpha$) &  $-$150 to $-$45\,\AA{} \\
Li abundance & $-$3.0$\pm$0.4 dex  \\
\hline
\label{tab_HyadesL5:param}
\end{tabular}
\end{table}
%

%
%%%%%%%%%%%%%%%%%%%%%%%%%%%%%%%%%%%%%%%%%%%
%%%%% Figure: Li vs Teff of Hyades members  %%%%%
%%%%%%%%%%%%%%%%%%%%%%%%%%%%%%%%%%%%%%%%%%%
%
\begin{figure}
  \includegraphics[width=0.95\linewidth, angle=0]{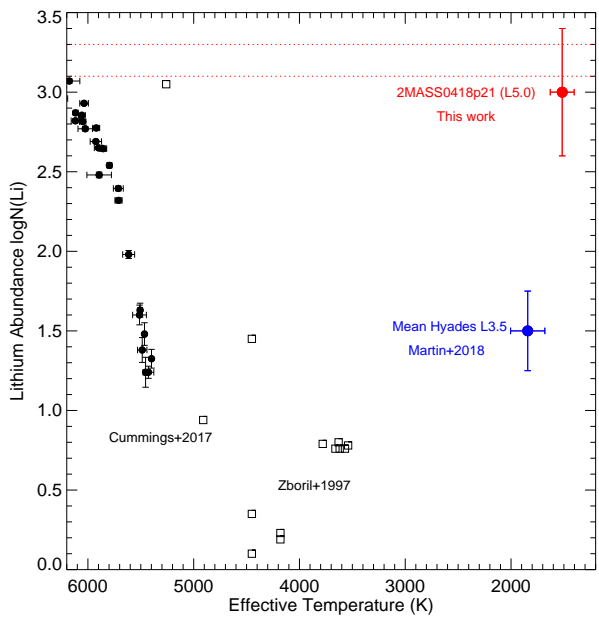}
  \caption{Lithium abundances (in the scale of $\log$\,N(H)\,=\,12) versus T$_{\rm eff}$ for Hyades 
solar-type stars \citep[black dots;][]{cummings17a}, K5--M4 stars \citep[upper limits;][]{zboril97a}, 
the mean value for L3.5 members \citep[blue dot;][]{martin18a}, and the BD 2M0418 (red dot).
The range of initial lithium meteoritic content is shown with red dotted lines as an upper
limit of the possible content in BD atmospheres.
}
  \label{fig_HyadesL5:Li_vs_Teff}
\end{figure}
%

%
%%%%%%%%%%%%%%%%%%%%%%%%%%%%%%%%%%%%%%%%%
%%%%% Figure: pEW Li vs SpT  %%%%%
%%%%%%%%%%%%%%%%%%%%%%%%%%%%%%%%%%%%%%%%%
%
\begin{figure}
 \centering
  \includegraphics[width=0.95\linewidth, angle=0]{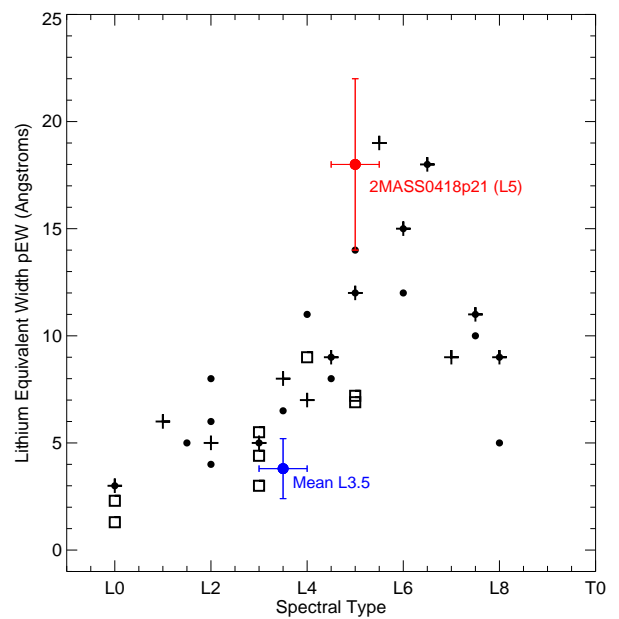}
  \caption{Lithium pseudo-equivalent widths (pEWs) of L and T dwarfs.
Measured lithium pEW for field \citep[dots and crosses;][]{kirkpatrick00,kirkpatrick08} 
and young L dwarfs \citep[open squares;][]{cruz09}.
2M0418 is highlighted with a red dot and the mean value for L3.5 members of the Hyades in blue.
}
   \label{fig_HyadesL5:pEW_SpT}
\end{figure}
%

%
%%%%%%%%%%%%%%%%%%%%%%%%%%%%%%
%%%%% New Li detection  %%%%%
%%%%%%%%%%%%%%%%%%%%%%%%%%%%%%
%
\section{New lithium detection in 2M0418}
\label{HyadesL5:newLi}

The lithium test was proposed in the 1990s to distinguish between low-mass
stars and BDs \citep{rebolo92}. The technique relies on the detection
of lithium in absorption at 6707.8\,\AA{} for late-M and L dwarfs straddling the
stellar to sub-stellar boundary. If the lithium is detected in an ultracool dwarf, the object is sub-stellar.
If not, it is a high-mass BD or a very low-mass star.

Figure \ref{fig_HyadesL5:Spectrum} shows the combined optical spectrum of 2M0418, calibrated in flux.
In the right-hand side panel we show the region of the optical
spectrum around the Li absorption including a significant part of the continuum to appreciate the detection.
We measured a pseudo-equivalent width (pEW) of 18$\pm$4\,\AA{} (1$\sigma$), consistent with the 
compilation of field L5 dwarfs exhibiting lithium in absorption 
We derived the error bars from the dispersion of the ten individual spectra of the target in two
continuum regions around the lithium line applying Equation 7 of \citet{cayrel88a}.
\citep{kirkpatrick99,kirkpatrick00,kirkpatrick08,pavlenko07,cruz09,zapatero14a,lodieu15b}.
In Fig.\ \ref{fig_HyadesL5:Spectrum} we over-plot in red the spectrum of a field L5V,
2MASS\,J12392727$+$5515371, exhibiting one of the strongest lithium absorption among mid-L dwarfs 
and a decent signal-to-noise ratio in the region of lithium. We infer a pEW of 15$\pm$2\AA{} and a 
full-width-half-maximum of 26$\pm$4\AA{}, in agreement with the 14\AA{} reported by \citet{kirkpatrick99}.

The mass of 2M0418 was estimated by \citet{perez-garrido17} using the near/mid-IR colours reported in 
Table 1 of that paper and the evolutionary models of the Lyon group \citep{allard01,chabrier00c}. A likely 
mass between 39--55 $M_{\rm Jup}$ with central value of 48 $M_{\rm Jup}$ was inferred.
Our new detection of lithium confirms the sub-stellarity of 2M0418 with an upper limit of 
$\sim$60 $M_{\rm Jup}$ (see Section \ref{HyadesL5:age_Hyades}).

Figure \ref{fig_HyadesL5:Li_vs_Teff} shows the evolution of the lithium pEWs as a function of
T$_{\rm eff}$ with a minimum around K and M dwarfs due to the burning of lithium.
\citet{cummings17a} gives direct A(Li)=\,$\log$\,(N(Li)/N(H)) measurements for solar-type members of the Hyades.
We measured upper limits of about 0.1\AA{} on the pEWs of the lithium in the optical 
spectra published in \citet{stauffer97} and kindly provided by the author. These values translate into
lithium abundances below $\log$\,N(Li)\,=\,1.0 based on the work by \citet{zboril97a} which quote 
equivalent widths of a few \AA{} for field K5--M6 dwarfs equivalent to abundances around the plateau at 0.85\@.
We added 2M0418 in Fig.\ \ref{fig_HyadesL5:Li_vs_Teff} for which we 
inferred the abundance as follows. We investigated the curve of growth created from the
BT-Settl atmospheric structure computed by \citet{allard12} and the theoretical code WITA2 of 
\citet{pavlenko97a} and \citet{pavlenko07} for gravities of $\log$(g)\,=\,4.5 and 5.0 dex and
T$_{\rm eff}$ of 1600K, 1800K, 2000K, 2100K, and 2200K (green dots in 
Fig.\ \ref{fig_HyadesL5:Li_vs_Teff}).
The closest predicted pEW of lithium at 6707.8\AA{} is given by $\log$(g)\,=\,5.0 dex and
T$_{\rm eff}$\,=\,1600\,K (17.9\AA{}), yielding an abundance of $\log$\,N(Li)\,=\,3.0$\pm$0.4 dex.
The uncertainty of 0.4 dex on the abundance comes mainly from the error on the pEW\@.
We also added to the plot the mean abundance derived from the mean pEW of lithium derived for two 
Hyades L3.5 members \citep{martin18a}.

%
%%%%%%%%%%%%%%%%%%%%%%%%%%%%%%%%%%%
%%%%% Persistent Halpha %%%%%
%%%%%%%%%%%%%%%%%%%%%%%%%%%%%%%%%%%
%
\section{H$\alpha$ emission}
\label{HyadesL5:Halpha}

Since the first optical spectrum collected on 25 January 2015 with GTC/OSIRIS \citep{perez-garrido17},
we see persistent but variable emission in H$\alpha$ at 6562.8\,\AA{}. We measured the pEW on each of the 
ten individual spectra and found a variation 
in the strength of the H$\alpha$ line from $-$150\,\AA{} to $-$45\,\AA{} over the period of approximately 
one month (23 January till 28 February 2017).

%
%%%%%%%%%%%%%%%%%%%%%%%%%%%%%%%%%%%%%%%%%%%
%%%%% Figure: Teff (or Lum) vs Age %%%%%
%%%%%%%%%%%%%%%%%%%%%%%%%%%%%%%%%%%%%%%%%%%
%
\begin{figure}
  \includegraphics[width=0.95\linewidth, angle=0]{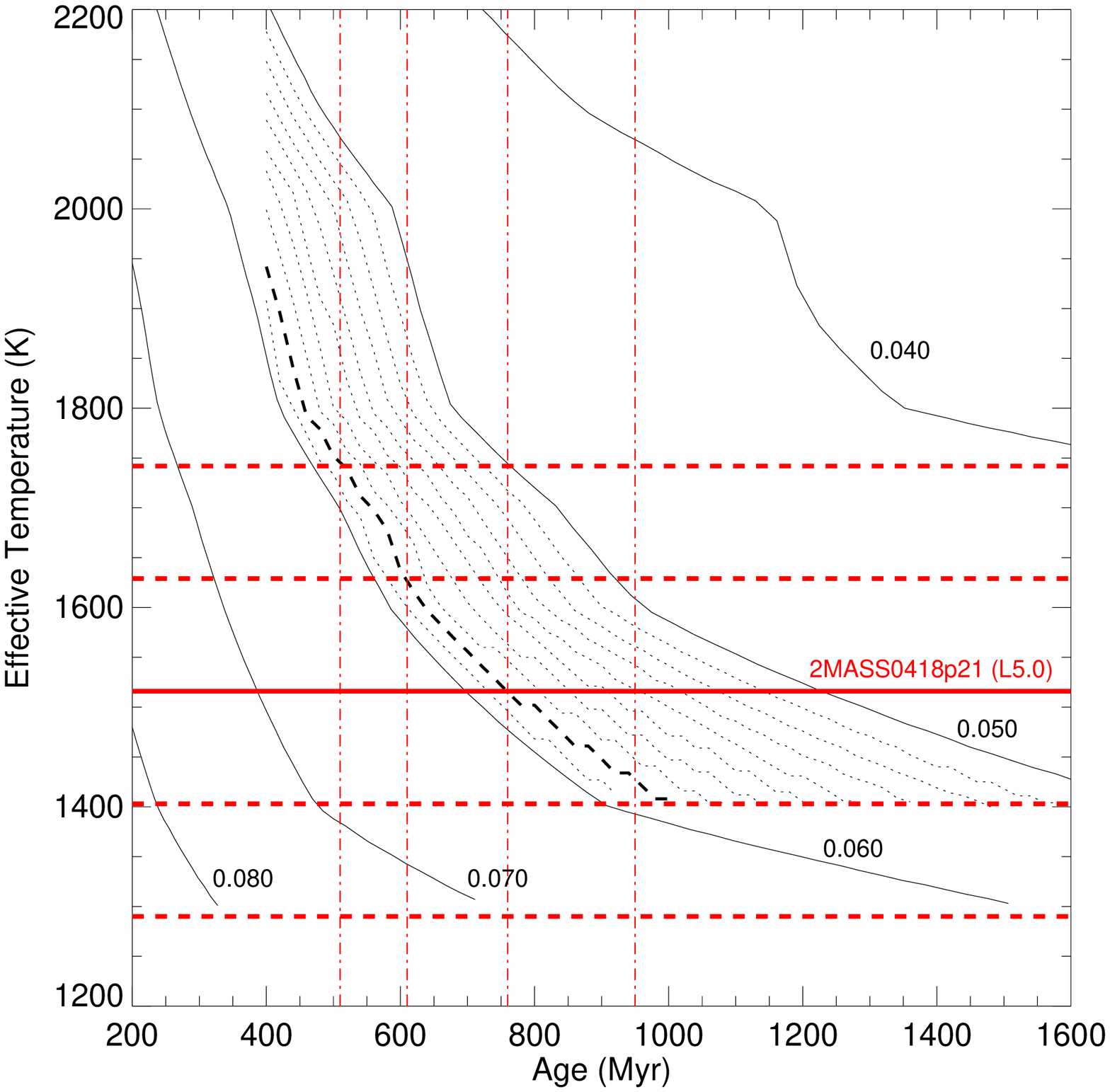}
  \includegraphics[width=0.95\linewidth, angle=0]{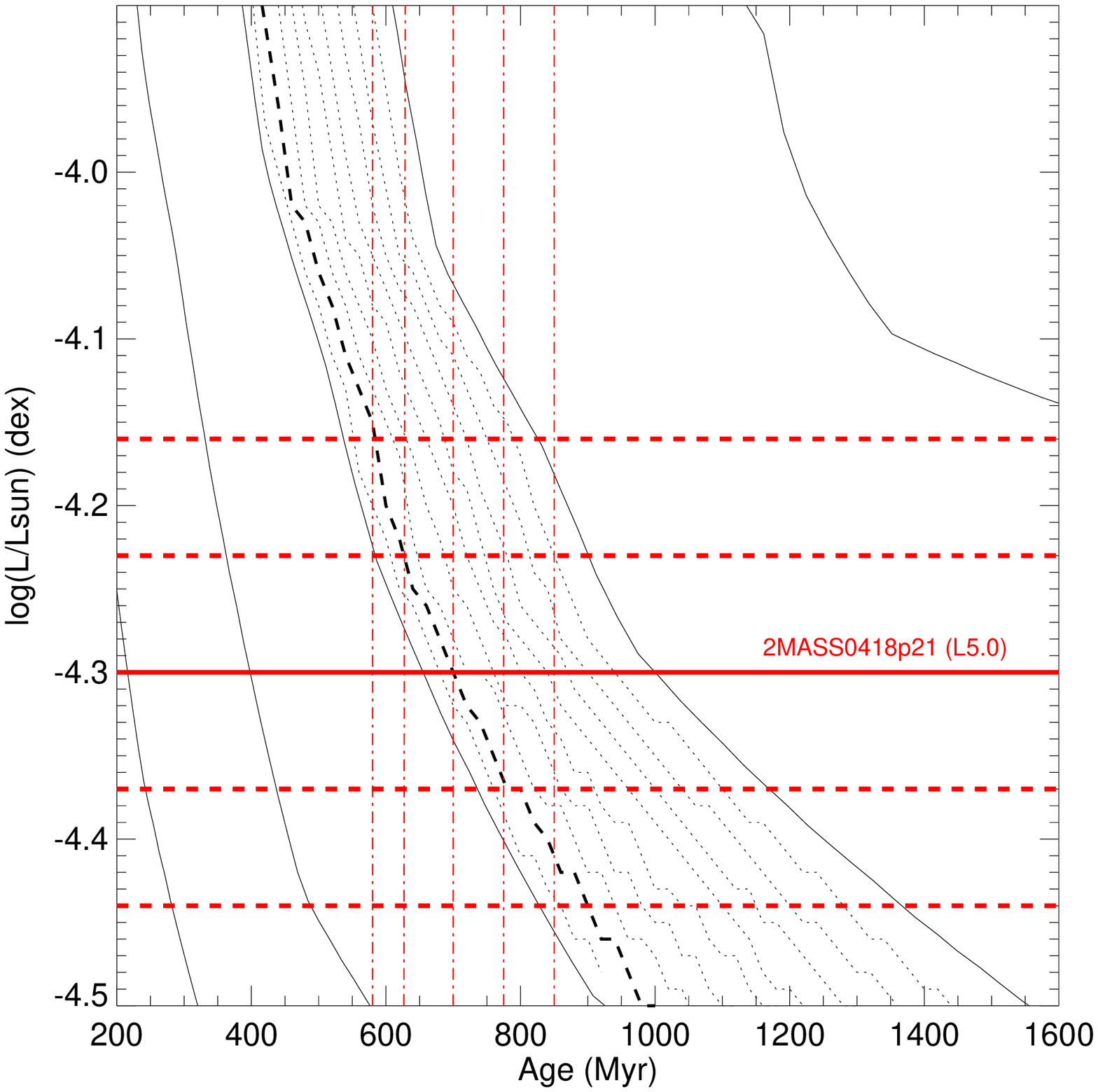}
  \caption{{\it{Top:}} T$_{\rm eff}$ vs age of isomasses from the BHCA95 models
for masses of 10--80 $M_{\rm Jup}$ by steps of 1 $M_{\rm Jup}$ (black lines). The 51--59
$M_{\rm Jup}$ isomasses are highlighted with black dotted lines for a limited age range (400--1400 Myr). 
{\it{Bottom:}} Luminosity versus age.
2M0418 is marked with a red line and the 1 and 2$\sigma$ error bars shown as dashed lines.
}
  \label{fig_HyadesL5:Teff_vs_Age}
\end{figure}

%
%%%%%%%%%%%%%%%%%%%%%%%%%%%%%%%%%%%
%%%%% Age of the Hyades %%%%%
%%%%%%%%%%%%%%%%%%%%%%%%%%%%%%%%%%%
%
\section{Age of the Hyades cluster}
\label{HyadesL5:age_Hyades}

In this section we discuss the constraints that we can set on the age of the Hyades cluster based on 
the presence of lithium in absorption in the optical spectrum of 2M0418\@. 

The study of chondritic meteorites suggests an original lithium elemental abundance A(Li)\,=\,3.28$\pm$0.05 
\citep{lodders03} in the solar system. We assumed a similar elemental abundance in the Hyades cluster. 
In Fig.\ \ref{fig_HyadesL5:Li_vs_Teff} we observe that the maximum abundance of solar-type members 
of the Hyades lies around 3.0 dex, very similar to the lithium abundance of the interstellar 
medium \citep{ferlet84a,lemoine93} and the Pleiades \citep{boesgaard88a,garcia_lopez94a}.
Based on the upper envelope of L dwarfs with clear lithium absorption lines,
2M0418 might have fully preserved its lithium because the largest pEW in field L dwarfs
is 15\AA{} (filled dots in Fig.\ \ref{fig_HyadesL5:pEW_SpT}) for 
2MASSW\,J1239272$+$551537 \citep{kirkpatrick00,jameson08a} resolved into a 0.21$''$ binary
\citep{gizis03}. Our lithium abundance computations were implemented for 1D model atmospheres under
the assumption of local thermodynamic and hydro-static equilibrium with no sinks of energy. The chemical 
equilibrium was computed for $\approx$100 molecular species with WITA2 \citep{pavlenko97a,pavlenko07,pavlenko13}.

The detection of lithium in 2M0418 sets a upper limit on its mass and on the age of the Hyades.
We studied the preservation of lithium predicted by the latest BHAC15 models \citep{baraffe15} for
masses below 60 $M_{\rm Jup}$ (black lines in Fig.\ \ref{fig_HyadesL5:Teff_vs_Age}). Due to the limited 
steps in masses, we requested to I.\ Baraffe the prediction of lithium for ages between 400 and 1.4 Gyr
and masses in the 50--60 $M_{\rm Jup}$ range. The BHAC15 models predict ratios of lithium to meteoritic 
lithium of $\sim$93--95\%, 89--91\%, and 81--85\% for masses of 50, 51, and 52 $M_{\rm Jup}$,
respectively (black dotted lines in Fig.\ \ref{fig_HyadesL5:Teff_vs_Age}). 
The evolution of this ratio is very quick, implying that the upper mass of 2M0418 is constrained to 
52 $M_{\rm Jup}$ assuming levels of depletion smaller than 20\%. However, we caution that this upper limit
depends on models and should be validated in the future with dynamical masses of mid-L dwarf binary
members of the Hyades. Therefore, the 1$\sigma$ upper limit on the age of the Hyades 
is set to 775 and 950 Myr from the luminosity and T$_{\rm eff}$ of 2M0418, respectively 
(Fig.\ \ref{fig_HyadesL5:Teff_vs_Age}).

Using a similar study of four L3--L4 photometric, astrometric, and radial velocity members of the Hyades, 
\citet{martin18a} infer a lower limit of 580 Myr
on the age of the cluster, which combined with our upper limit would correspond to a range of 580--775 Myr
(mean of 700 Myr) and 580--950 Myr (mean of 760 Myr) at 1$\sigma$ for the Hyades from the luminosity and 
T$_{\rm eff}$ of 2M0418\@.
This age range is in agreement with the original estimate of 625$\pm$50 Myr from \citet{maeder81}
and the recent lithium depletion age of \citet{martin18a} but on the lower end of the 750$\pm$100 Myr
determination \citep{brandt15a,brandt15b}.
We note the importance of identifying a large sample of L dwarfs to refine the age of the
cluster and impose further constraints on the physics of the interior of BDs.
The Hyades is the oldest open cluster with an age derived by the lithium depletion boundary technique.

%
%%%%%%%%%%%%%%%%%%%%%%%%%%%%%%%
%%%%%  ACKNOWLEDGEMENTS %%%%%
%%%%%%%%%%%%%%%%%%%%%%%%%%%%%%%
%
\begin{acknowledgements}
We thank the referee for a detailed report that improved the quality of this manuscript.
This research has been supported by the Spanish Ministry of Economy and Competitiveness (MINECO) 
under the grants AYA2015-69350-C3-2-P and AYA2015-69350-C3-3-P\@.
We thank Yakiv Pavlenko for his calculations of equivalent widths at different temperatures, Isabelle
Baraffe for her models, and Eduardo Mart\'in for sharing his results prior to publication. 

This work is based on observations made with the Gran Telescopio Canarias (GTC), 
operated on the island of La Palma in the Spanish Observatorio del Roque de los 
Muchachos of the Instituto de Astrof\'isica de Canarias (programme GTC77-16B led by P\'erez Garrido).

This research has made use of the Simbad and Vizier databases, operated
at the Centre de Donn\'ees Astronomiques de Strasbourg (CDS), and
of NASA's Astrophysics Data System Bibliographic Services (ADS).

We thank John Stauffer for kindly providing the optical spectra of the K and M Hyades members,
published in 1997 \citep{stauffer97b}.
\end{acknowledgements}
%

%
%%%%%%%%%%%%%%%%%%%%%%%%%%%%%%%%%%%%%%%%%%
%%%%%%%%  Bibliography  %%%%%%%%
%%%%%%%%%%%%%%%%%%%%%%%%%%%%%%%%%%%%%%%%%%
%
%\begin{thebibliography}{}
\bibliographystyle{aa}
\bibliography{../../AA/mnemonic,../../AA/biblio_old} 

\end{document}